\input{aipcheck}

\documentclass[
    ,final            
  ]
  {aipproc}

\usepackage{lineno}
\layoutstyle{6x9}

\begin{document}

\title{Multiwavelength analysis of four millisecond pulsars}


\author{L. Guillemot}{
  address={Max-Planck-Institut f\"ur Radioastronomie, Auf dem H\"ugel 69, 53121 Bonn, Germany, \texttt{guillemo@mpifr-bonn.mpg.de}}
}

\author{I. Cognard}{
  address={Laboratoire de Physique et Chimie de l'Environnement, LPCE UMR 6115 CNRS, F-45071 Orl\'eans Cedex 02, France}
}

\author{T.~J.~Johnson}{
  address={University of Maryland, Departments of Physics and Astronomy, College Park, MD 20742, USA}
}

\author{C.~Venter}{
  address={Centre for Space Research, North-West University, Potchefstroom 2520, South Africa}
}

\author{A.~K.~Harding}{
  address={NASA Goddard Space Flight Center, Greenbelt, MD 20771, USA}
}

\author{the \emph{Fermi} LAT Collaboration, Pulsar Timing Consortium and Pulsar Search Consortium}{
address={Across the world}
}

\begin{abstract}
Radio timing observations of millisecond pulsars (MSPs) in support of \emph{Fermi} LAT observations of the gamma-ray sky enhance the sensitivity of high-energy pulsation searches. With contemporaneous ephemerides we have detected gamma-ray pulsations from PSR B1937+21, the first MSP ever discovered, and B1957+20, the first known black-widow system. The two MSPs share a number of properties: they are energetic and distant compared to other gamma-ray MSPs, and both of them exhibit aligned radio and gamma-ray emission peaks, indicating co-located emission regions in the outer magnetosphere of the pulsars. However, radio observations are also crucial for revealing MSPs in \emph{Fermi} unassociated sources. In a search for radio pulsations at the position of such unassociated sources, the Nan\c cay Radio Telescope discovered two MSPs, PSRs J2017+0603 and J2302+4442, increasing the sample of known Galactic disk MSPs. Subsequent radio timing observations led to the detection of gamma-ray pulsations from these two MSPs as well. We describe multiwavelength timing and spectral analysis of these four pulsars, and the modeling of their gamma-ray light curves in the context of theoretical models.
\end{abstract}

\maketitle


\section{Introduction}

In contrast with normal pulsars, most millisecond pulsars (MSPs) have binary orbits, making them extremely difficult to find in blind searches of the gamma-ray data alone, because of the vast parameter space to be searched and the sparsity of the gamma-ray photons. Gamma-ray pulsation searches from MSPs therefore rely heavily on observations at radio frequencies, with observations lasting only a few minutes to several hours. For known MSPs, searches for pulsations are done by folding the high-energy data using astrometric, rotational and orbital parameters measured through radio timing observations. Fourteen MSPs have been observed to emit pulsed gamma rays with this strategy in the first two years of data recorded by the \emph{Fermi} Large Area Telescope (LAT) \cite{Fermi8MSPs,FermiJ0034,Romani2010}. Another approach consists in searching \emph{Fermi} unassociated sources for radio pulsations from unknown pulsars, a strategy which has so far led to more than 20 discoveries of MSPs at the Green Bank, Parkes, Nan\c cay and Effelsberg telescopes \cite{Romani2010}. 

The famous MSPs J1939+2134 (a.k.a. B1937+21) and J1959+2048 (B1957+20), respectively the first millisecond pulsar ever discovered \cite{Backer1982} and the first known black-widow system \cite{Fruchter1988}, are examples of MSPs observed to emit pulsed gamma rays by folding the data using timing parameters measured at the Nan\c cay and Westerbork telescopes \cite{Guillemot2011}. The Nan\c cay Radio Telescope has also contributed to the wealth of discoveries of new MSPs in \emph{Fermi} unassociated sources by finding the two previously unknown MSPs J2017+0603 and J2302+4442 \cite{Cognard2011}.

\section{Two MSPs with aligned radio and gamma-ray emission}

All MSPs so far detected with the \emph{Fermi} LAT in gamma rays have spin-down energy loss rates $\dot E$ above 10$^{33}$ erg s$^{-1}$, making J1939+2134 ($\dot E = 1.1 \times 10^{36}$ erg s$^{-1}$) and J1959+2048 ($\dot E = 7.5 \times 10^{34}$ erg s$^{-1}$) likely gamma-ray emitters. Nevertheless, the two MSPs are located at low Galactic latitude, and hence suffer from intense contamination by the diffuse gamma-ray emission in the Galaxy. In addition, both have relatively large distances of $d = (7.7 \pm 3.8)$ kpc for J1939+2134 \cite{Verbiest2009} and $d = (2.5 \pm 1.0)$ kpc for J1959+2048 \cite{NE2001}, so that their ``spin-down luminosities'' $\dot E / d^2$ are on the low end of gamma-ray MSPs detected thus far with \emph{Fermi}. 

\begin{figure}
  \includegraphics[scale=0.33]{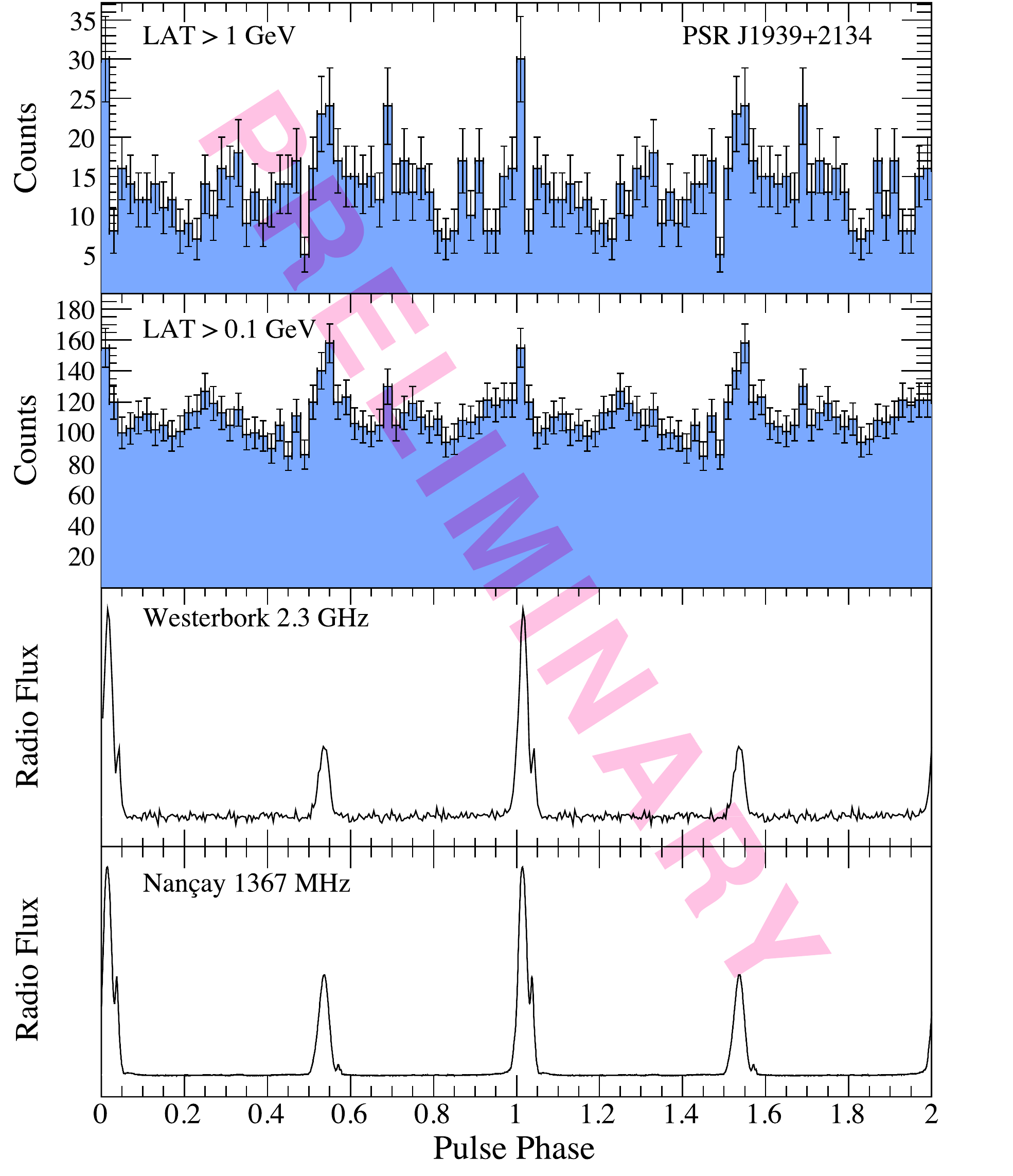}
  \includegraphics[scale=0.33]{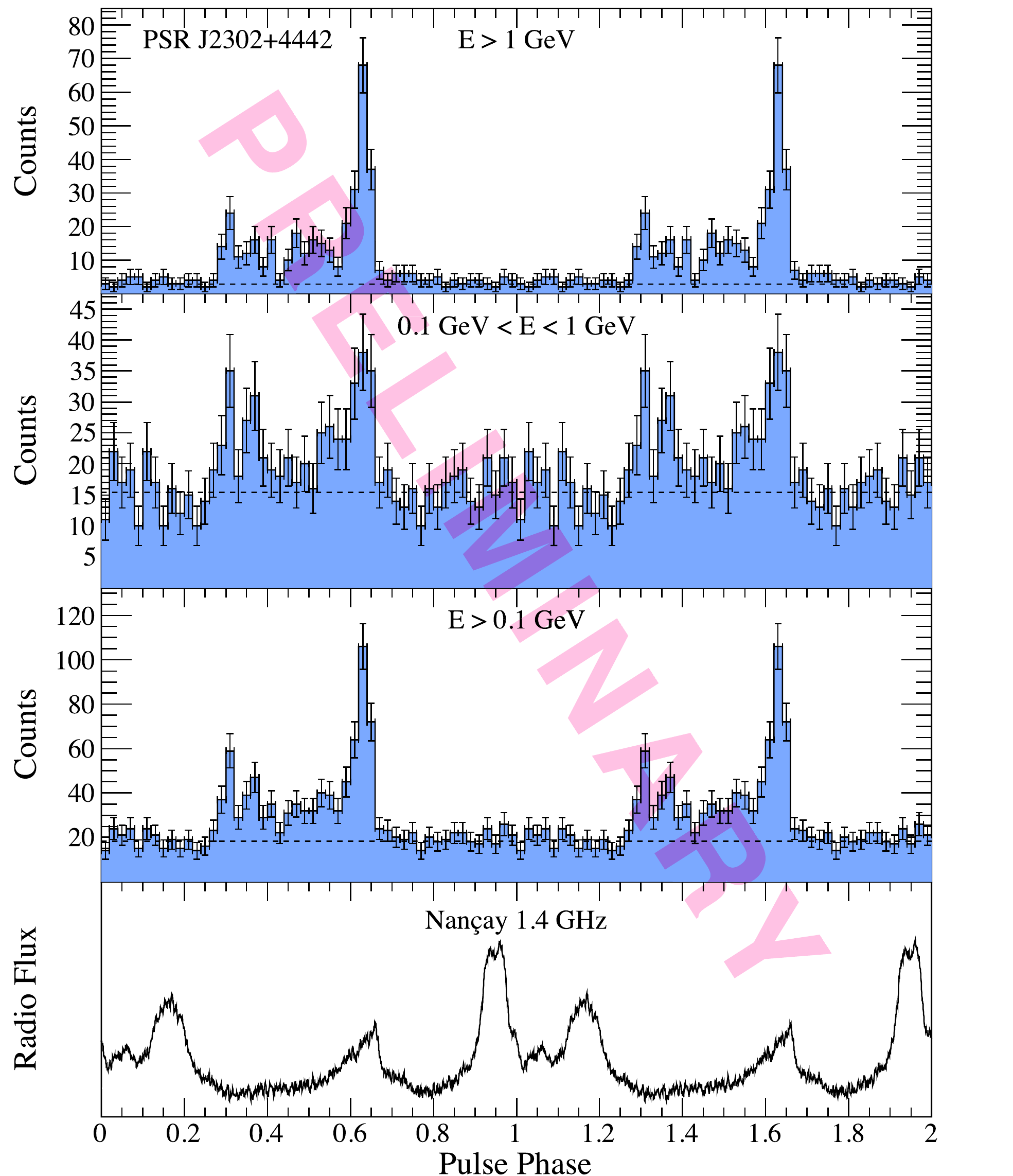}

  \caption{Radio and gamma-ray light curves of PSRs J1939+2134 and J2302+4442. Two pulsar rotations are shown for clarity. See \cite{Guillemot2011,Cognard2011} for details on the analysis of the radio and gamma-ray data.  \label{lc1939}}
\end{figure}

\begin{figure}
  \includegraphics[scale=0.8]{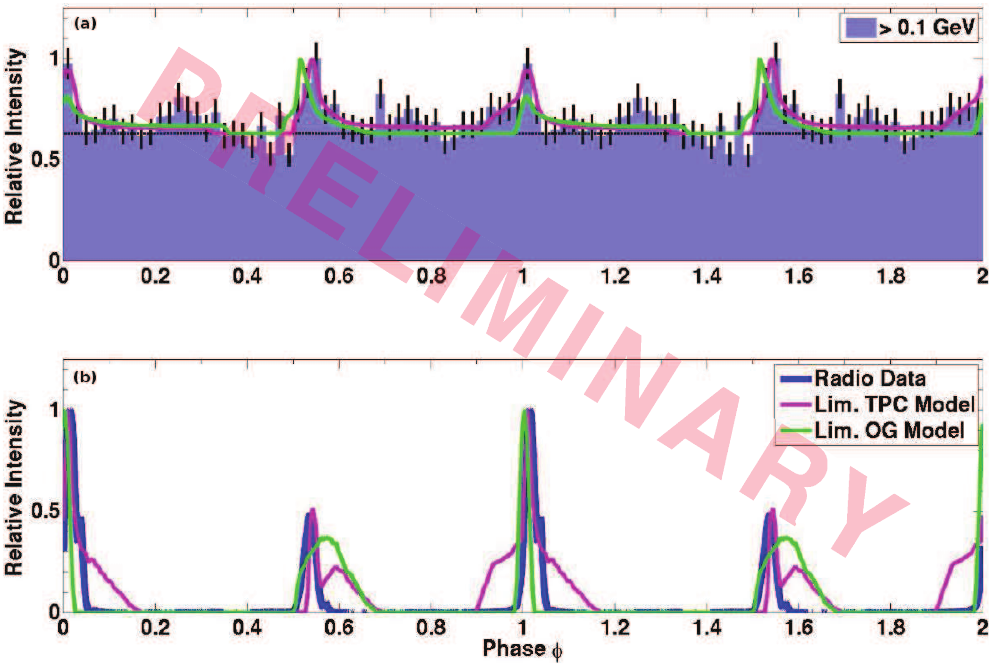}
  \caption{Model fits for the radio and gamma-ray light curves of PSR J1939+2134.\label{modeling1939}}
\end{figure}

Both MSPs have been monitored at radio wavelengths in the context of the timing campaign of best gamma-ray candidates for detection by \emph{Fermi} \cite{Smith2008}. We phase-folded the first 18 months of gamma-ray data recorded by the LAT using pulsar ephemerides based on radio timing data taken at the Nan\c cay and Westerbork radio telescopes, and detected pulsed gamma-ray emission from J1939+2134 and J1959+2048. Figure \ref{lc1939} shows radio and gamma-ray light curves of J1939+2134. The close alignment of radio and gamma-ray emission peaks, also observed for PSR J1959+2048 and J0034$-$0534 \cite{FermiJ0034}, suggests that for this subset of gamma-ray MSPs low and high-energy radiation is produced in the same regions of the magnetosphere. As illustrated in Figure \ref{modeling1939}, radio and gamma-ray profiles of PSRs J1939+2134 and J1959+2048 are well reproduced by ``altitude-limited'' Two-Pole Caustic (TPC) and Outer Gap (OG) models, in which both radio and gamma-ray emission peaks result from caustics generated at high altitude with respect to the light cylinder \cite{Venter2011}, confirming that radio and gamma-ray emissions seem co-located for these MSPs. 

\section{New MSPs discovered at Nan\c cay}

Of the 1451 gamma-ray sources listed in the \emph{Fermi} Large Area Telescope First Catalog (1FGL) \cite{Fermi1FGL}, 630 are not associated with counterparts known at other wavelengths. Sources of gamma-ray emission with no known counterparts and with emission properties that resemble those of gamma-ray pulsars (lack of flux variation over time, spectral cutoff at 1 -- 10 GeV) can be searched for pulsations from radio pulsars. Observations of \emph{Fermi} unassociated sources likely to be pulsars conducted at Nan\c cay yielded the discovery of two MSPs in binary orbits, PSRs J2017+0603 ($P = 2.896$ ms) and J2302+4442 ($P = 5.192$ ms). Because the two MSPs were found in unassociated gamma-ray sources, they were likely to emit gamma rays in a pulsed way. We verified this assumption by folding the gamma-ray data using ephemerides based on timing data collected at the Nan\c cay, Jodrell Bank, and GBT telescopes, and detected pulsations in the gamma-ray data recorded after the discoveries. We then used the entire \emph{Fermi} dataset to extend the timing solutions to the past, thereby improving the measurement of astrometric and rotational parameters. Detailed timing parameters for the two new MSPs are given in \cite{Cognard2011}.

Figure \ref{lc1939} shows an integrated radio profile at 1.4 GHz and gamma-ray light curves of PSR J2302+4442. These light curves are reminiscent of those observed for other gamma-ray MSPs, with relatively complex radio profiles and two gamma-ray peaks offset from the radio emission. Detailed modeling of the observed radio and gamma-ray light curves using standard outer magnetospheric models shows that, similarly to other gamma-ray MSPs, the gamma-ray emission seems to originate at high altitudes in their magnetospheres (see \cite{Johnson2010} for details on the light curve modeling of PSR J2017+0603). 

\section{Conclusion}

Most of the $>$20 MSPs newly discovered in unassociated gamma-ray sources are expected to be pulsed gamma-ray emitters, so that more than a third of currently known gamma-ray pulsars are MSPs, making them a prominent class of Galactic sources of gamma rays. The four MSPs discussed here illustrate the importance of radio observations in support of the \emph{Fermi} mission, such as pulsar timing measurements and pulsar searches conducted at the Nan\c cay radio telescope. The Nan\c cay observatory will continue to perform timing observations of MSPs for \emph{Fermi} and search for new pulsars in unassociated sources. A third MSP, PSR J2043+1711 ($P = 2.380$ ms) has been discovered recently in a \emph{Fermi} source with no counterpart, and will hopefully be followed by others. 


\begin{theacknowledgments}

The $Fermi$ LAT Collaboration acknowledges support from a number of agencies and institutes for both development and the operation of the LAT as well as scientific data analysis. These include NASA and DOE in the United States, CEA/Irfu and IN2P3/CNRS in France, ASI and INFN in Italy, MEXT, KEK, and JAXA in Japan, and the K.~A.~Wallenberg Foundation, the Swedish Research Council and the National Space Board in Sweden. Additional support from INAF in Italy and CNES in France for science analysis during the operations phase is also gratefully acknowledged. The Nan\c cay Radio Observatory is operated by the Paris Observatory, associated with the French Centre National de la Recherche Scientifique (CNRS). The Westerbork Synthesis Radio Telescope is operated by Netherlands Foundation for Radio Astronomy, ASTRON. 


\end{theacknowledgments}


\bibliographystyle{aipproc}   

\end{document}